\documentclass[11pt]{article}

\usepackage[utf8]{inputenc}
\usepackage{graphicx}
\usepackage{color}
\usepackage{amsmath}
\usepackage{amssymb}
\usepackage{amsthm}  
\usepackage[colorlinks=true,linkcolor=black,citecolor=black]{hyperref}
\usepackage{palatino}

\newtheorem{thm}{Theorem}
\newtheorem{defin}{Definition}

\newcommand{\bra}[1]{\langle #1|}
\newcommand{\ket}[1]{|#1 \rangle}

\newcommand{\ketbra}[1]{\ket{#1}\bra{#1}}

\newcommand{\ident}{\mathbb{I}}
\DeclareMathOperator{\tr}{\mathrm{Tr}}

\newcommand{\mdag}{^{\dag}} % dag operator

\newcommand{\demi}{\frac{1}{2}}

\DeclareMathOperator{\In}{in}
\DeclareMathOperator{\Out}{out}
\DeclareMathOperator{\Typ}{typ}
\newcommand{\aeq}{\approx_{(a)}}

\begin{document}
\title{Coding for quantum channels with side information at the transmitter}
\author{
Frédéric Dupuis\\[1mm]
{\it\small Département d'informatique et de recherche opérationnelle, Université de Montréal}\\
{\it\small School of Computer Science, McGill University}\\
{\tt\small dupuisf@iro.umontreal.ca}
}
%\email{dupuisf@iro.umontreal.ca}
%\affiliation{Département d'informatique et de recherche opérationnelle, Université de Montréal}
%\affiliation{School of Computer Science, McGill University}

\maketitle

\begin{abstract}
We consider the problem of coding for quantum channels with side information that is available ahead of time at the transmitter but not at the receiver. We find a single-letter expression for the entanglement-assisted quantum capacity of such channels which closely parallels Gel'fand and Pinsker's solution to the classical version of the same problem. This theorem can also be used to find a lower bound on the unassisted quantum capacity of these channels.
\end{abstract}

\section{Introduction}
Consider the following problem: we have a noisy quantum memory device that can store $n$ qubits and which contains a certain fraction of defective cells. The cells that do work can be modelled as a depolarizing channel, but the defective ones always output $\ket{0}$. We can test which cells are defective before writing to the memory device, but this information is not necessarily available when reading from it. What is the best asymptotic rate at which we can store qubits reliably on this device? This problem can be generalized to any channel where the transmitter has access to side information about the channel state while the receiver does not.

The corresponding classical problem has been solved by Gel'fand and Pinsker in \cite{gelfand-pinsker}. They consider channels modelled as a conditional probability distribution $p_{Y|XS}(y|x,s)$, $x \in \mathcal{X}, s \in \mathcal{S}, y \in \mathcal{Y}$, where $x$, $y$ and $s$ represent the input, output and state of the channel respectively. The channel state is i.i.d. and distributed according to $p_S(s)$. The encoder has access to the entire sequence of channel states ahead of time whereas the decoder does not. They have shown that the capacity of such a channel is given by
\begin{equation}
	C = \max_{q_{USX} \in \mathcal{P}} \left[ I(U;Y) - I(U;S) \right]
	\label{eqn:gp-classique}
\end{equation}
where $\mathcal{P}$ is the set of all probability distributions on $\mathcal{U} \times \mathcal{X} \times \mathcal{S}$ such that the marginal on $\mathcal{S}$ is equal to $p_S(s)$; $\mathcal{U}$ is an arbitrary set that can be chosen such that $|\mathcal{U}| \leqslant |\mathcal{X}| + |\mathcal{S}|$. The mutual informations are computed on the distribution $p_{Y|XS} \cdot q_{USX}$.

Here we shall generalize this result to quantum channels. Namely, we will prove that the entanglement-assisted quantum capacity of quantum channels with side information at the transmitter has the same form as (\ref{eqn:gp-classique}) and, a relatively rare fact in quantum information theory, has a single-letter converse.

Many other quantum information processing tasks involving side information or multiple users have been considered in the literature. For instance, data compression with side information at the receiver (generally known as Slepian-Wolf coding \cite{slepian-wolf}) was generalized to the quantum world in \cite{devetak-winter-spcoding} in the case when only the side information is quantum, and in \cite{FQSW} when both the data and the side information are quantum. A slightly different generalization of Slepian-Wolf coding called \emph{state merging} was presented in \cite{merging}; in this case, the side information and the data to be sent are both quantum, but transmission is achieved through entanglement and classical communication. Classical channel simulation with quantum side information at the receiver is considered in \cite{luo-devetak}. Another primitive called \emph{quantum state redistribution} \cite{redistribution} allows one to send the $C$ subsystem of a mixed state $\rho^{ABC}$ from Alice (who initially holds both $A$ and $C$) to Bob who already holds $B$. Quantum generalizations of broadcast channels \cite{broadcast,bcast} and multiple access channels \cite{hdw05,winter-qmac,klimovitch-qmac} have also been considered.

Our result is very much in the same spirit as those in \cite{mother-father}, \cite{merging}, and \cite{FQSW}. In \cite{mother-father}, it is shown that a large number of quantum information protocols, such as one-way entanglement distillation \cite{DW05}, entanglement-assisted channel coding \cite{lsd1,lsd2,lsd3}, channel simulation \cite{reverse-shannon} and many others can be derived by simple transformations from two basic protocols, called the \emph{mother} and \emph{father} protocols. In \cite{merging}, the authors analyze the effect of making random measurements on a state and show that this leads to the state merging protocol, which can be used to derive several additional protocols. In \cite{FQSW}, a fully quantum version of state merging is presented, called the ``fully quantum Slepian-Wolf'' protocol (FQSW) from which all other protocols mentioned (including the father, the mother, state merging, and state redistribution \cite{oppenheim-redist-proof}) can be derived. This generality is not surprising: the state merging and FQSW papers essentially consist of an analysis of the power of randomly selected unitary transformations, which can be viewed as the quantum generalization of random codes. We will use this approach here for our direct coding theorem, replacing Gel'fand and Pinsker's random binning argument by a random unitary, as in the FQSW theorem.

In the next section, we will introduce our notation, define precisely what a quantum channel with side information at the transmitter is, and give appropriate definitions of achievable rate and capacity. Section \ref{sec:fqsw} will consist of a quick review of the FQSW protocol. Section \ref{sec:direct} will be devoted to the direct coding theorem itself, with the single-letter converse given in section \ref{sec:converse}. 

\section{Notation, definitions and background material}
Quantum subsystems will be denoted by capital letters $A$, $B$, and so on; we will use superscripts on states to indicate which subsystems they are defined on. Given a ket $\ket{\psi}$, we will use the shorthand $\psi = \ketbra{\psi}$ for its associated density operator. For any density operator, we will denote its partial trace on one subsystem by removing that subsystem from the superscript, i.e. $\rho^{A} = \tr_B \rho^{AB}$. Furthermore, we will use the notation $A^n$ to denote the tensor product of $n$ copies of the system $A$.

For operations on quantum states, including unitaries, isometries and superoperators, we use a similar convention to denote the input and output spaces; for example, $\mathcal{N}^{A \rightarrow B}$ will denote a superoperator from $A$ to $B$. Superscripts will be omitted when doing so is not likely to cause confusion. The dimension of a system $A$ is denoted $|A|$.

A superoperator can always be extended to an isometry by adding another output subsystem representing the environment. The operation performed by this extension is exactly the same as the original channel if we trace out the environment system. We will denote the isometric extension of $\mathcal{N}^{A' \rightarrow B}$ by $U_{\mathcal{N}}^{A' \rightarrow BE}$. Here $U_{\mathcal{N}}$ does not act on density operators but on the Hilbert space itself.

To make the notation less cumbersome, we will use the symbol $\cdot$ to denote conjugation; i.e. $A \cdot B := ABA\mdag$.

We will also define $\ket{\Phi}^{SS'} := \frac{1}{\sqrt{|S|}} \sum_{i=0}^{|S|} \ket{ii}^{SS'}$, where the $\ket{i}^S$ and $\ket{i}^{S'}$ are some standard orthonormal bases on $S$ and $S'$. 

We will often use the trace norm, which we define to be $\|M\|_1 := \tr|M|$ for any Hermitian matrix $M$. This norm induces a metric $\|\rho-\sigma\|_1$ on the space of quantum states.

The von Neumann entropy of a density operator $\rho^A$ will be denoted $H(A)_{\rho}$; when $\rho$ is defined on more than one subsystem, $H(A)_{\rho}$ denotes the entropy of its restriction to $A$. The quantum mutual information is the function $I(A;B)_{\rho} = H(A)_{\rho} + H(B)_{\rho} - H(AB)_{\rho}$, where $\rho$ is some state on systems $A$ and $B$ (and possibly more subsystems).

We will say that two families of states $\psi$ and $\varphi$ parametrized by their size $n$ are asymptotically equal (denoted $\psi \aeq \varphi$) if $\|\psi - \varphi\|_1$ goes to zero as $n \rightarrow \infty$. See Appendix \ref{sec:asymptotic} for a formal definition.

Finally, we will often make use of a version of Uhlmann's theorem \cite{uhlmann} shown in \cite{DHW05}, lemma 2.2: whenever $\psi^A$ and $\varphi^A$ are density matrices such that $\|\psi^A - \varphi^A\|_1 \leqslant \varepsilon$, for any pair of purifications $\psi^{AB}$ and $\varphi^{AC}$ of $\psi^A$ and $\varphi^A$ respectively, there exists a partial isometry $V^{C \rightarrow B}$ such that $\|\psi^{AB} - V \cdot \varphi^{AC} \|_1 \leqslant 2\sqrt{\varepsilon}$. In particular, whenever we have two families of states $\psi^A$ and $\varphi^A$ such that $\psi^A \aeq \varphi^A$, then for any family of purifications $\psi^{AB}$ and $\varphi^{AC}$ there exists a family of partial isometries $V^{C\rightarrow B}$ such that $\psi^{AB} \aeq V \cdot \varphi^{AC}$.

\subsection{Definition of quantum channels with side information at the transmitter}

A quantum channel with side information at the transmitter is defined by a superoperator $\mathcal{N}^{A'S \rightarrow B}$ and a quantum state $\ket{\psi}^{SS'}$; this quantum state represents the side information. Alice has access to $S'$ and can input a state of her choice into $A'$. One way to view this is to say that Alice shares entanglement with the channel itself. This framework allows us to consider both quantum and classical side information about the channel in a unified manner.

To illustrate this, consider the example of the depolarizing channel with defects given in the introduction. For this case, we can choose $\ket{\psi}$ to be $\sqrt{p} \ket{00} + \sqrt{1-p} \ket{11}$. The superoperator $\mathcal{N}$ then measures the $S$ subsystem, and outputs $\ket{0}$ if the outcome is $0$. If the outcome is $1$, it applies the depolarizing channel to $A'$ and sends the output to Bob.

In this paper, we will be mostly interested in the entanglement-assisted quantum capacity of such channels.

We will now define precisely the notions of code, achievable rate, and capacity for quantum channels with side information at the transmitter. Even though the main concern of the paper is the entanglement-assisted capacity, we are nonetheless interested in the amount of entanglement consumed by our protocol. We will therefore consider protocols in which some of the entanglement used during the execution of the protocol is returned at the end, since doing this improves the entanglement consumption rate of the protocol.

\begin{defin}\label{def:code}
	A $(Q,n,\varepsilon)$-code for a quantum channel with side information at the transmitter $(\mathcal{N},\ket{\psi})$ consists of an encoding superoperator $\mathcal{E}^{R' \tilde{A} S'^{n} \rightarrow A'^{n} \hat{A}}$ and a decoding superoperator $\mathcal{D}^{B^{n} \tilde{B} \rightarrow \bar{B} \hat{B}}$ such that
	\begin{equation}
		\left\| \mathcal{D}(\mathcal{N}(\mathcal{E}(\varphi_{\In}))) - \varphi_{\Out} \right\|_1 \leqslant \varepsilon
	\end{equation}
	where $\varphi_{\In} = \Phi^{RR'} \otimes \Phi^{\tilde{A}\tilde{B}} \otimes (\psi^{SS'})^{n}$, $\varphi_{\Out} = \Phi^{R \bar{B}} \otimes \Phi^{\hat{A} \hat{B}}$, and $\log |R| = \log|R'| = nQ$.
\end{defin}
It will turn out to be much more convenient for us to use isometric extensions of the encoding and decoding superoperators. We will generally use $W^{R' \tilde{A} S'^n \rightarrow A'^n \hat{A} D}$ to denote the isometric extension of the encoding map $\mathcal{E}$ and $V^{B^{n} \tilde{B} \rightarrow \bar{B} \hat{B} G}$ for the decoding map $\mathcal{D}$.

\begin{defin}\label{def:achievable-rate}
	We say that $Q$ is an \emph{achievable rate} for the channel $(\mathcal{N}, \ket{\psi})$ if there exists a sequence of $(Q,n,\varepsilon_n)$-codes such that $\varepsilon_n \rightarrow 0$ as $n \rightarrow \infty$.
\end{defin}
In other words, $Q$ is achievable if there exists a family of codes as defined above such that
\begin{equation}
	V \cdot U_{\mathcal{N}}^{\otimes n} \cdot W \cdot \varphi_{\In} \aeq \varphi_{\Out}
\end{equation}

The capacity of a channel $(\mathcal{N},\ket{\psi})$ is the supremum of all achievable rates.

The goal of this paper is to establish the following theorem:
\begin{thm}\label{thm:main}
	The entanglement-assisted quantum capacity of a quantum channel with side information at the transmitter $( \mathcal{N}, \ket{\psi})$ is
	\begin{equation}\label{eqn:cap}
		C = \sup_{\sigma} \left\{ \demi I(A;B)_{\omega} - \demi I(A;S)_{\sigma} \right\} 
	\end{equation}
	The supremum is taken over all mixed states of the form $\sigma^{AA'S}$ where $\sigma^S = \psi^S$ and $\omega = \mathcal{N}(\sigma)$.
\end{thm}
Of course, this theorem entails that the entanglement-assisted classical capacity of quantum channels with side information at the transmitter is
\begin{equation}
	C = \sup_{\sigma} \left\{ I(A;B)_{\mathcal{N}(\sigma)} - I(A;S)_{\sigma} \right\}
\end{equation}
via super-dense coding.

\section{The FQSW theorem}\label{sec:fqsw}
Before presenting our protocol, we first give a quick overview of the fully quantum Slepian-Wolf protocol \cite{FQSW}. Suppose Alice and Bob hold a mixed state $\rho^{AB}$. We introduce a reference system $R$ to purify the state; the resulting state is $\ket{\psi}^{ABR}$. Alice would like to transfer her state to Bob with very high fidelity by sending him as few qubits as possible. The FQSW theorem states that Alice can do this by first applying a unitary transformation to her entire share of the state (a random unitary selected according to the Haar measure will work with high probability), splitting her share into two subsystems $\bar{A}$ and $\hat{A}$, and then sending $\hat{A}$ to Bob.

Note that this scheme works provided that the subsystems $\bar{A}$ and $R$ are in a product state after applying the random unitary: since Bob holds the purifying system of $\bar{A}R$, there exists a local unitary that Bob can apply to turn his purifying system into separate purifying systems of the two subsystems. The purifying system of $R$ is exactly the original state that Alice wanted to send to Bob together with the share Bob originally had, and $\bar{A}$ together with its purifying system is an EPR pair shared by Alice and Bob. This last feature is an added bonus of the protocol: Alice and Bob get some free entanglement at the end.

It is possible to calculate how close $\bar{A}$ and $R$ are to being in a product state. The result of the calculation is the following (see \cite{FQSW} for details):
\begin{equation}
    \int_{\mathbb{U}(A)} \left\| \rho^{\bar{A} R}(U) - \frac{\ident^{\bar{A}}}{|\bar{A}|} \otimes \psi^R \right\|_1^2 dU \leqslant \frac{|A||R|}{|\hat{A}|^2} \tr\left[ \left( \psi^{AR} \right)^2 \right]
    \label{eqn:fqsw}
\end{equation}
where $\rho_{\bar{A}R}(U) = \tr_{\hat{A}}[U \cdot \psi^{AR}]$. Since the inequality holds for the average over choices of $U$, there must exist at least one $U$ that satisfies it.

A special case of interest is when the initial state is an i.i.d. state of the form $(\ket{\psi}^{ABR})^{\otimes n}$. In this case, it can be shown that as long as $\log |\hat{A}| \geqslant n[\frac{1}{2}I(A;R) + \delta]$, it will be true that
\begin{equation}\label{eqn:asymptotic-fqsw}
\varphi^{\bar{A}R^{\otimes n}} \aeq \frac{\ident^{\bar{A}}}{|\bar{A}|} \otimes \varphi^{R^{\otimes n}}
\end{equation}
where $\varphi^{\bar{A}\hat{A}B^{\otimes n}R^{\otimes n}}$ is the result of applying the random unitary to $\Pi_A \cdot (\psi^{ABR})^{\otimes n}$, where $\Pi_A$ is the projector onto the typical subspace of the $A$ subsystem, as defined in Appendix \ref{sec:typical}, and $\delta > 0$.

Note that it is also possible to show that the value $\left\| \rho^{\bar{A} R}(U) - \frac{\ident^{\bar{A}}}{|\hat{A}|} \otimes \psi^R \right\|_1^2$ is exponentially concentrated around its mean value. Hence, by the union bound, given a constant number of equations of the form (\ref{eqn:asymptotic-fqsw}) in which the random unitary is applied to the same system, for $n$ large enough there must exist a single family of random unitaries that satisfies all of them at the same time. We will make use of this fact in the direct coding theorem.

\section{Direct coding theorem}\label{sec:direct}
The direct coding theorem is very similar to the one in \cite{bcast}. We start out with $n$ copies of $\sigma^{AA'S}$ from theorem \ref{thm:main} and construct the $n$th term in a sequence of codes achieving a rate of $\tfrac{1}{2}[I(A;B)_{\mathcal{N}(\sigma)} - I(A;S)_{\sigma}]$. It will be convenient for us to purify $\sigma$ by introducing an additional subsystem $D$. 

%\begin{figure}
%	\centering
%	\input{states-diagram.pdf_t}
%	\caption{Diagram illustrating the relationships between the various states defined in section \ref{sec:direct}. Solid lines indicate exact equality, whereas dotted lines indicate asymptotic equality.}
%	\label{fig:states-diagram}
%\end{figure}

The way to derive a code from $\sigma^{\otimes n}$ is to transform this state into one which looks like $\varphi_{\In}$ in definition \ref{def:code}. To do this, we will first restrict $\sigma^{\otimes n}$ to its typical subspace on $A^n$ using a family of typical projectors $\Pi_A^{A^n \rightarrow A_{\Typ}}$ (see Appendix \ref{sec:typical}); we then have $nH(A)_{\sigma} - n\delta_n \leqslant \log|A_{\Typ}| \leqslant nH(A)_{\sigma} + n\delta_n$, with $\delta_n \rightarrow 0$ as $n \rightarrow \infty$. Then, for each $n$, we shall apply a random unitary on $A_{\Typ}$ and split it into three subsystems: $\tilde{B}$, $\hat{A}$, and $R$, with $\frac{n}{2} I(A;ED)_{\mathcal{N}(\sigma)} + 2n\delta_n$ qubits, $\frac{n}{2} I(A;S)_{\sigma} + 2n\delta_n$ qubits, and $\log|A_{\Typ}| - \log|\tilde{B}| - \log|\hat{A}|$ qubits respectively (hence $\log|R| \geqslant \frac{n}{2} I(A;B)_{\mathcal{N}(\sigma)} - \frac{n}{2} I(A;S)_{\sigma} - 5n\delta_n$). Call this random unitary $U^{A_{\Typ} \rightarrow R \tilde{B} \hat{A}}$. According to the FQSW theorem, with high probability over the choice of $U$, both of the following will hold:

\begin{align}
	(U \cdot \Pi_A \cdot \sigma^{\otimes n})^{R \tilde{B} S^n} &\aeq \frac{\ident^{R\tilde{B}}}{|R\tilde{B}|} \otimes (\psi^S)^{\otimes n} \label{eqn:RBtildeS-decoupled}\\
	(U_{\mathcal{N}}^{\otimes n} \cdot U \cdot \Pi_A \cdot \sigma^{\otimes n})^{R \hat{A} E^nD^n} &\aeq \frac{\ident^{R\hat{A}}}{|R\hat{A}|} \otimes (\mathcal{N}(\sigma)^{ED})^{\otimes n} \label{eqn:RAhatED-decoupled}
\end{align}

By Uhlmann's theorem, this implies that there exist families of partial isometries $W^{R' \tilde{A} S'^n \rightarrow A'^n \hat{A} D^n}$ and $V^{\tilde{B} B^n \rightarrow \bar{B} \hat{B} G}$ (one $W$ and one $V$ for each $n$) and a family of pure states $\xi^{G E^n D^n}$ such that

\begin{align}
	\label{eqn:condition-W}
	(U \cdot \Pi_A \cdot \sigma^{\otimes n})^{R \tilde{B} S^n A'^n D^n} &\aeq W \cdot \Phi^{RR'} \otimes \Phi^{\tilde{B}\tilde{A}} \otimes (\psi^{SS'})^{\otimes n}\\ 
	(V \cdot U_{\mathcal{N}}^{\otimes n} \cdot U \cdot \Pi_A \cdot \sigma^{\otimes n})^{R \hat{A} \tilde{B} B^n E^nD^n} &\aeq \Phi^{R\bar{B}} \otimes \Phi^{\hat{A}\hat{B}} \otimes \xi^{GE^nD^n}
	\label{eqn:condition-V}
\end{align}
Here, (\ref{eqn:condition-W}) follows from the fact that the left-hand side (resp. right-hand side) of (\ref{eqn:condition-W}) is a purification of the left-hand side (resp. right-hand side) of (\ref{eqn:RBtildeS-decoupled}); $W$ is the unitary required by Uhlmann's theorem to make the two sides close in trace distance. Equation (\ref{eqn:condition-V}) follows from (\ref{eqn:RAhatED-decoupled}) in a similar manner. Combining these two equations and using the fact that asymptotic equality is transitive, we get that
\begin{equation}
	V \cdot U_{\mathcal{N}}^{\otimes n} \cdot W \cdot  \left(\Phi^{RR'} \otimes \Phi^{\tilde{B}\tilde{A}} \otimes (\psi^{SS'})^{\otimes n}\right) \aeq \Phi^{R\bar{B}} \otimes \Phi^{\hat{A}\hat{B}} \otimes \xi^{GE^nD^n}
	\label{eqn:victory-condition}
\end{equation}
which proves the direct coding theorem.

\subsection{Entanglement cost and unassisted transmission}
One can check the rate at which this protocol consumes entanglement between the transmitter and the receiver. The size of $\tilde{B}$ is $\frac{n}{2} I(A;ED)_{\mathcal{N}(\sigma)} + 2n\delta_n$ qubits; however, one can see that some entanglement is recovered at the end of the protocol in the $\hat{A}$ and $\hat{B}$ subsystems, with $\log|\hat{A}| = \frac{n}{2} I(A;S)_{\sigma}+2n\delta_n$. Hence, the net amount of entanglement consumed is $\frac{n}{2}\left[ I(A;ED)_{\mathcal{N}(\sigma)} - I(A;S)_{\sigma} \right]$.

Furthermore, this allows us to calculate the transmission rate when this protocol is used to send qubits without preshared entanglement. In this mode, we use part of the transmission rate to send the entanglement needed by the protocol, and the rest of the rate to send the data we are actually interested in sending. We compute this rate by subtracting the entanglement cost from the number of transmitted qubits: $\log|R| - \log|\tilde{B}| + \log|\hat{A}| \geqslant \frac{n}{2}I(A;B)_{\mathcal{N}(\sigma)} - \frac{n}{2}I(A;S)_{\sigma} - \frac{n}{2}I(A;ED)_{\mathcal{N}(\sigma)} + \frac{n}{2}I(A;S)_{\sigma} - 5n\delta_n = \frac{n}{2}I(A\rangle B)_{\mathcal{N}(\sigma)} - 5n\delta_n$. We therefore have a rate of $\frac{1}{2} I(A\rangle B)_{\mathcal{N}(\sigma)}$ for unassisted transmission. This expression is of the same form as for the usual unassisted quantum coding theorem, although the coherent information is defined on a slightly different state.

\section{Converse theorem}\label{sec:converse}
We shall now prove that for any achievable rate $Q$, there exists a state $\sigma^{AA'S}$ as in theorem \ref{thm:main} for which $Q = \demi I(A;B)_{\mathcal{N}(\sigma)} - \demi I(A;S)_{\sigma}$. This part is essentially the same as in \cite{gelfand-pinsker}, with a few adaptations to the quantum case. In particular, one must pay close attention to which state the various mutual informations are defined on, since we will be dealing with states where only some fraction of the $n$ instances of the channel has been applied.

First, let $W^{\tilde{A} R' S'^n \rightarrow A'^n \hat{A}}$ and $V^{\tilde{B}B^n \rightarrow \bar{B} \hat{B} G}$ be encoding and decoding isometries for a $(Q,n,\varepsilon)$-code as in definition \ref{def:code}, and let $\varphi_{\In} = \Phi^{RR'} \otimes \Phi^{\tilde{A}\tilde{B}} \otimes (\psi^{SS'})^{\otimes n}$, $\sigma = W \cdot \varphi_{\In}$ and $\omega = U_{\mathcal{N}}^{\otimes n} \cdot \sigma$. Then, by Fannes' inequality we must have that
\begin{equation}
	I(R;B^{n} \tilde{B})_{\omega} \geqslant 2n(Q - d(\varepsilon,n))
\end{equation}
where $d(\varepsilon,n) := \frac{3\varepsilon Q}{2} + \frac{3\varepsilon \log \varepsilon}{n}$.  Notice that
\begin{align}
	I(R;\tilde{B} B^{n})_{\omega} &= I(\tilde{B};R)_{\omega} + I(R;B^{n}|\tilde{B})_{\omega}\\
	&= I(R;B^{n}|\tilde{B})_{\omega} \label{eqn:weak-converse-2}\\
	&\leqslant I(R\tilde{B};B^{n})_{\omega}
\end{align}
where (\ref{eqn:weak-converse-2}) is due to the fact that $R$ and $\tilde{B}$ are independent. Combining this with $I(R\tilde{B};S^{n})_{\sigma} = 0$, we have
\begin{equation}
	I(R\tilde{B};B^{n})_{\omega} - I(R\tilde{B};S^{n})_{\sigma} \geqslant 2n(Q - d(\varepsilon,n))
\end{equation}
We will now introduce a few shorthands which will make the notation considerably less cumbersome: we will write $B^i$ instead of $B_1,\ldots,B_i$ and $B_i^j$ instead of $B_i,\ldots,B_j$, and likewise for $S$. Define also
\begin{align}
	X(i) &:= R\tilde{B} B^{i-1} S_{i+1}^n\\
	Y(i) &:= R\tilde{B} S_{i+1}^n
	\label{eqn:uv-defs}
\end{align}
Note that these are nothing more than groupings of subsystems. We also define the following sequence of states:
\begin{equation}
	\omega(i) := (U_{\mathcal{N}}^{\otimes i} \otimes \ident^{\otimes n-i}) \cdot \sigma
	\label{eqn:sigma-omega-defs}
\end{equation}
In other words, $\omega(i)$ is the result of applying the first $i$ instances of the channel to the state $\sigma$.

We shall now prove the inequality
\begin{equation}
	I(R\tilde{B};B^{n})_{\omega} - I(R\tilde{B};S^{n})_{\sigma} \leqslant \sum_{i=1}^n \left\{ I(X(i);B_i)_{\omega(i)} - I(X(i);S_i)_{\omega(i-1)} \right\}.
	\label{eqn:toprove}
\end{equation}
Since each term in this sum is of the form $I(A;B)_{\mathcal{N}(\sigma)} - I(A;S)_{\sigma}$ for some $\sigma^{AA'S}$, the highest term is achievable by the direct coding theorem and therefore there exists a state for which $Q \leqslant I(A;B)_{\mathcal{N}(\sigma)} - I(A;S)_{\sigma}$. This allows us to conclude the theorem.

We now proceed in exactly the same way as in \cite{gelfand-pinsker} to establish (\ref{eqn:toprove}): we consider the inequality
\begin{multline}
	I(Y(i);B^i)_{\omega(i)} - I(Y(i);S^i)_{\omega(i-1)} \leqslant \left[ I(Y(i-1);B^{i-1})_{\omega(i-1)} - I(Y(i-1);S^{i-1})_{\omega(i-2)} \right]\\
	+ \left[ I(X(i);B_i)_{\omega(i)} - I(X(i);S_i)_{\omega(i-1)} \right].
	\label{eqn:eq17}
\end{multline}
Summing up all these inequalities from $i=2$ to $i=n$, we obtain (\ref{eqn:toprove}), since $Y(n) = R\tilde{B}$ and $Y(1) = X(1)$.

Now, to prove (\ref{eqn:eq17}), we use the following identities which follow from the definitions of $X(i)$ and $Y(i)$ and from basic properties of the mutual information.

\begin{align}
	I(Y(i);B^i)_{\omega(i)} &= I(Y(i);B^{i-1})_{\omega(i)} + I(Y(i);B_i|B^{i-1})_{\omega(i)}\\
	I(Y(i);S^i)_{\omega(i-1)} &= I(Y(i);S_i)_{\omega(i-1)} + I(Y(i);S^{i-1}|S_i)_{\omega(i-1)}\\
	I(Y(i-1);S^{i-1})_{\omega(i-1)} &= I(Y(i);S^{i-1}|S_i)_{\omega(i-1)}\\
	I(Y(i-1);B^{i-1})_{\omega(i-1)} &= I(S_i;B^{i-1})_{\omega(i-1)} + I(Y(i);B^{i-1}|S_i)_{\omega(i-1)}\\
	I(X(i);B_i)_{\omega(i)} &= I(B^{i-1};B_i)_{\omega(i)} + I(Y(i); B_i|B^{i-1})_{\omega(i)}\\
	I(X(i);S_i)_{\omega(i-1)} &= I(B^{i-1};S_i)_{\omega(i-1)} + I(Y(i);S_i|B^{i-1})_{\omega(i-1)}
\end{align}

Substituting these into (\ref{eqn:eq17}) and using the identity
\begin{equation}
	I(A;B) - I(A;B|C) = I(A;C) - I(A;C|B)
	\label{eqn:mutual-info-identity}
\end{equation}
which holds on any mixed state $\rho^{ABC}$, we get that the difference between the right-hand side and the left-hand side of (\ref{eqn:eq17}) is $I(B^{i-1};B_i)_{\omega(i)}$, which is always nonnegative. This concludes the proof.

\section{Discussion and Conclusion}\label{sec:discussion}
This result further strengthens the parallel between classical information theory problems and their entanglement-assisted quantum counterparts. Indeed, the capacity formula (\ref{eqn:cap}) has the same form as the classical version (\ref{eqn:gp-classique}); the same phenomenon arises in the case of the entanglement-assisted capacities of regular point-to-point channels \cite{BSST02}, multiple-access channels \cite{hdw05}, and, for the best coding theorem we know, broadcast channels \cite{bcast}. A similar equivalence has also been shown at the level of error-correcting codes themselves \cite{hsieh-devetak-brun1, hsieh-devetak-brun2, hsieh-devetak-brun3}. It would be particularly interesting to have a systematic way in which classical coding theorems could be transformed into entanglement-assisted quantum protocols as it would enable us to import much larger classes of results from classical information theory into the quantum world.

Returning to our result, there is one remaining issue that one would like to solve in order to have a fully satisfactory characterization of the achievable rate region: we currently have no upper bound on the dimension of the $A$ system needed to achieve the capacity in expression (\ref{eqn:cap}). Thus, despite having a single-letter converse, we unfortunately do not have a way to compute the capacity. In the classical case, it is possible to use Carathéodory's theorem to bound the cardinality of $\mathcal{U}$ in the optimal input distribution. However, in the quantum case, this approach fails due to the fact that the quantum conditional entropy cannot in general be expressed as $H(A|B) = \sum_b p(b) H(A|B=b)$. On the other hand, there is little reason to believe that large dimensions are necessary to achieve the optimal rate, but we have not managed to prove it. In fact, one encounters a very similar difficulty when trying to calculate the squashed entanglement \cite{squashed-ent} of a particular state since we have no bound on the size of the subsystem we need to condition on. We therefore leave this issue as an open problem.

One might also wonder about a related problem: whether the capacity can in general be achieved by optimizing only over pure states $\sigma^{AA'S}$. This would imply an upper bound on $|A|$. However, one can show that this cannot be the case: take, for example, a qubit-to-qubit channel which applies one of the four Pauli operations with equal probability, but where $S$ tells the transmitter which one of the four operations is applied. The capacity of such a channel is clearly one qubit per transmission. Suppose that this rate is achievable using a pure state $\sigma^{AA'S}$. Then, we must have $\demi I(A;B)_{\mathcal{N}(\sigma)} = 1$ (since $B$ is two-dimensional) and therefore $\demi I(A;S)_{\sigma} = 0$. However, this last equation together with the fact that $\sigma$ is pure implies that the purification of $S$ must be entirely in $A'$. This is impossible since $S$ is maximally mixed over a four-dimensional system whereas $A'$ is two-dimensional, and hence the optimal $\sigma$ cannot be pure.

\section*{Acknowledgments}
The author would like to thank Patrick Hayden for many useful comments on a preliminary version of this paper.

\appendix

\section{Asymptotic equality}\label{sec:asymptotic}
Here we formally define the asymptotic equations involving the $\aeq$ relation. Let $\psi = \left\{ \psi_{(1)},\psi_{(2),\cdots} \right\}$ and $\varphi = \left\{ \varphi_{(1)},\varphi_{(2),\cdots} \right\}$ be two families of quantum states, where $\psi_{(n)}$ and $\varphi_{(n)}$ are defined on a Hilbert space $\mathcal{H}^{\otimes n}$. Then we say that $\psi \aeq \varphi$ if $\lim_{n \rightarrow \infty} \left\| \psi_{(n)} - \varphi_{(n)} \right\|_1 = 0$. We then say that $\psi$ and $\varphi$ are asymptotically equal. Note that, by the triangle inequality, $\aeq$ is transitive for any finite number of steps independent of $n$.

It should be mentioned that throughout the paper, asymptotic families of states are not always explicitly referred to as such, but generally speaking, whenever a state depends on the number of copies, it should be considered as a family of states. In addition, with a slight abuse of notation, we allow quantum operations on families of states; it should be clear which operation is done on each member of the family.

\section{Typical subspaces}\label{sec:typical}
Much of information theory relies on the concept of typical sequences. Let $\mathcal{X}$ be some alphabet and let $X$ be a random variable defined on $\mathcal{X}$ and distributed according to $p(x)$. Define the $\varepsilon$-typical set as follows:
\begin{equation*}
   \mathcal{T}_{\varepsilon}^{(n)} = \left\{ x^n \in \mathcal{X}^n \Big| \left|{-\tfrac{1}{n}}\log \Pr\{X^n = x^n\} - H(X) \right| \leqslant \varepsilon \right\}
%\mathcal{T}_{\varepsilon}^{(n)} = \left\{ x^n \in \mathcal{X}^n \Big| (\forall a \in \mathcal{X}) \left| N(a|x^n) - \Pr\{X=a\} \right| \leq \varepsilon / |\mathcal{X}| \right\}
\end{equation*}
where $X^n$ refers to $n$ independent, identically-distributed copies of $X$. It can be shown that the two following properties hold:
\begin{enumerate}
    \item There exists a function $\varepsilon(n)$ such that $\lim_{n \rightarrow \infty} \varepsilon(n) = 0$ and such that $\Pr\{X^n \in \mathcal{T}_{\varepsilon(n)}^{(n)}\} \geqslant 1-\varepsilon(n)$.
    \item There exists an $n_0$ such that for all $n>n_0$, $|\mathcal{T}_{\varepsilon}^{(n)}| \leqslant 2^{n[H(X) + \varepsilon]}$.
\end{enumerate}

The quantum generalization of these concepts is relatively straightforward: let $\rho^A = \sum_{x \in \mathcal{X}} p(x) \ketbra{x}$ be the spectral decomposition of a quantum state $\rho^A$ on a quantum system $A$. Then we can define the typical projector on the quantum system $A^{\otimes n}$ as follows:
\begin{equation*}
    \Pi_{\varepsilon}^{(n)} = \sum_{x^n \in \mathcal{T}_{\varepsilon}^{(n)}} \ketbra{x^n}
\end{equation*}
We call the support of $\Pi_{\varepsilon}^{(n)}$ the $\varepsilon$-typical subspace of $A^{\otimes n}$. (For brevity, we often omit $\varepsilon$ and refer simply to the typical subspace. In this case, unless otherwise stated, $\varepsilon$ can be assumed to be a positive constant, independent of $n$.) The two properties given above generalize to the quantum case:
\begin{enumerate}
    \item There exists a function $\varepsilon(n)$ such that $\lim_{n \rightarrow \infty} \varepsilon(n) = 0$ and such that $\tr\left[\Pi_{\varepsilon(n)}^{(n)} {\rho^A}^{\otimes n}\right] \geqslant 1-\varepsilon(n)$.
    \item There exists an $n_0$ such that for all $n>n_0$, $\tr[\Pi_{\varepsilon}^{(n)}] \leqslant 2^{n[H(A) + \varepsilon]}$.
\end{enumerate}

Note that the first of these two properties implies that $\Pi_{\varepsilon(n)}^{(n)} \cdot {\rho^A}^{\otimes n} \aeq {\rho^A}^{\otimes n}$, via the ``gentle measurement'' lemma (Lemma 9 in \cite{winter99}). With some abuse of notation, we will use $\Pi_A^{A^n \rightarrow A_{\Typ}}$ to refer to a family of typical projectors on $A^n$ which satisfies the two properties above.

\bibliographystyle{amsalpha}
\bibliography{biblio}

\end{document}